\documentclass[%
 reprint,
 amsmath,amssymb,
 aps,
floatfix,
]{revtex4-2}
\usepackage{graphicx}
\usepackage{dcolumn}
\usepackage{bm}
\usepackage{comment}
\usepackage{upgreek}
\usepackage{xcolor}
\usepackage{soul}

\begin{document}

\newcommand*\todo[1]{\colorbox{yellow}{#1}}

\preprint{APS/123-QED}

\title{Grain splitting is a mechanism for grain coarsening in colloidal polycrystals}

\author{Anna R. Barth}
\thanks{These authors contributed equally.}
 \affiliation{%
 Department of Physics, Harvey Mudd College, Claremont, CA
}%

\author{Maya H. Martinez}
\thanks{These authors contributed equally.}
 \affiliation{%
 Department of Physics, Harvey Mudd College, Claremont, CA
}%

\author{Cora E. Payne}
 \affiliation{%
 Department of Physics, Harvey Mudd College, Claremont, CA
}%

\author{Chris G. Couto}
 \affiliation{%
 Department of Physics, Harvey Mudd College, Claremont, CA
}%

\author{Izabela J. Quintas}
 \affiliation{%
 Department of Physics, Harvey Mudd College, Claremont, CA
}%

\author{Inq Soncharoen}
 \affiliation{%
 Department of Physics, Harvey Mudd College, Claremont, CA
}%

\author{Nina M. Brown}
 \affiliation{%
 Department of Physics, Harvey Mudd College, Claremont, CA
}%

\author{Eli J. Weissler}
 \affiliation{%
 Department of Physics, Harvey Mudd College, Claremont, CA
}%

\author{Sharon J. Gerbode}
 \email{gerbode@hmc.edu}
\affiliation{%
 Department of Physics, Harvey Mudd College, Claremont, CA
}%

\date{\today}

\begin{abstract}
In established theories of grain coarsening, grains disappear either by shrinking or by rotating as a rigid object to coalesce with an adjacent grain. Here we report a third mechanism for grain coarsening, in which a grain splits apart into two regions that rotate in opposite directions to match two adjacent grains' orientations. We experimentally observe both conventional grain rotation and grain splitting in 2D colloidal polycrystals. We find that grain splitting occurs via independently rotating ``granules'' whose shapes are determined by the underlying triangular lattices of the two merging crystal grains. These granules are so small that existing continuum theories of grain boundary energy are inapplicable, so we introduce a hard sphere model for the free energy of a colloidal polycrystal. We find that during splitting, the system overcomes a free energy barrier before ultimately reaching a lower free energy when splitting is complete. Using simulated splitting events and a simple scaling prediction, we find that the barrier to grain splitting decreases as grain size decreases. Consequently, grain splitting is likely to play an important role in polycrystals with small grains. This discovery suggests that mesoscale models of grain coarsening may offer better predictions in the nanocrystalline regime by including grain splitting.
\end{abstract} 


\maketitle

\section{\label{sec:Intro}Introduction}
The growth and merging of grains in polycrystalline materials, collectively called ``grain coarsening,'' plays a critical role in determining material properties \cite{Hansen2004, Jose2001, Li2020, Meyers2006, Huang2011, Hall1951, Watanabe1999}. For nearly a century, grain coarsening has been described by continuum theories in which the energetic cost of disordered grain boundaries creates a surface tension that drives grain boundary migration \cite{Read1950, GottsteinTextbook}. More recently, grain rotation has been postulated to play a significant role \cite{Erb1979, Trautt2012, Upmanyu2006}, particularly in nanocrystalline materials, where each grain contains only hundreds or thousands of atoms \cite{Haslam2001, Moldovan2002, Zhou2017, Chen2014,  Jin2004, Mompiou2014, Thomas2017, Ma2004, Shan2004, Wang2014, Wang2017, Noskova2007, moldovantextbook}. Various causes for grain rotation have been proposed, including shear coupling between neighboring grains \cite{Srinivasan2002, Cahn2004}, and a driving torque described by the Read-Shockley equation for the free energetic cost of a grain boundary \cite{Harris1998, Moldovan2001, Li1962, Martin1992}. Both classes of theories assume that grains rotate as rigid objects.

Colloidal polycrystals offer the opportunity to directly visualize grain coarsening at the particle scale. Previous groups have reported grain rotation of colloidal crystal grains that are only in contact with a single neighboring grain \cite{Lavergne2018, Moore2010, LiWei2020}. However, grain rotation in a colloidal polycrystal, where multiple adjacent grains can generate competing torques, has remained elusive.

Here we report experimental evidence of grain rotation in a colloidal polycrystal. Furthermore, despite the predictions of continuum theories that treat grains as rigidly rotating objects, we find that grains can also split apart into counterrotating regions. These regions are themselves composed of smaller ``granules'' that rotate independently. Using simulations of grain splitting events, we find that the free energetic cost of such grain splitting is prohibitively high for large grains, explaining why this phenomenon has been overlooked in continuum models of grain boundary motion. New models of grain coarsening that incorporate grain splitting may offer more accurate predictions for the structural dynamics of nanocrystalline materials.


\section{\label{sec:ExpMethods}Experimental methods: \protect\\ Preparing a simple polycrystal}

We prepare colloidal suspensions of silica spheres of diameter $1.3~\upmu$m (Sekisui Micropearl Spacers, Dana Enterprises International, CA),  in dimethyl sulfoxide. We pipette this suspension into a wedge-shaped cell constructed from two glass coverslides \cite{Gerbode2008}, and tilt the cell to allow the particles to sediment into the narrow end of the wedge, where they form an effectively 2D hard sphere crystalline monolayer. To create grain boundary loops within the monolayer, we use the ``optical blasting'' technique \cite{Cash2018}. Briefly, because the refractive index of the particles is less than that of the suspending fluid, a focused laser beam repels the particles. We use optical blasting to radially repel particles within the monolayer, creating space that attracts grain boundaries. This method can be used to move grain boundaries and create grains with custom shapes \cite{Cash2018}.

\begin{figure}[h!] \includegraphics[width=0.48\textwidth]{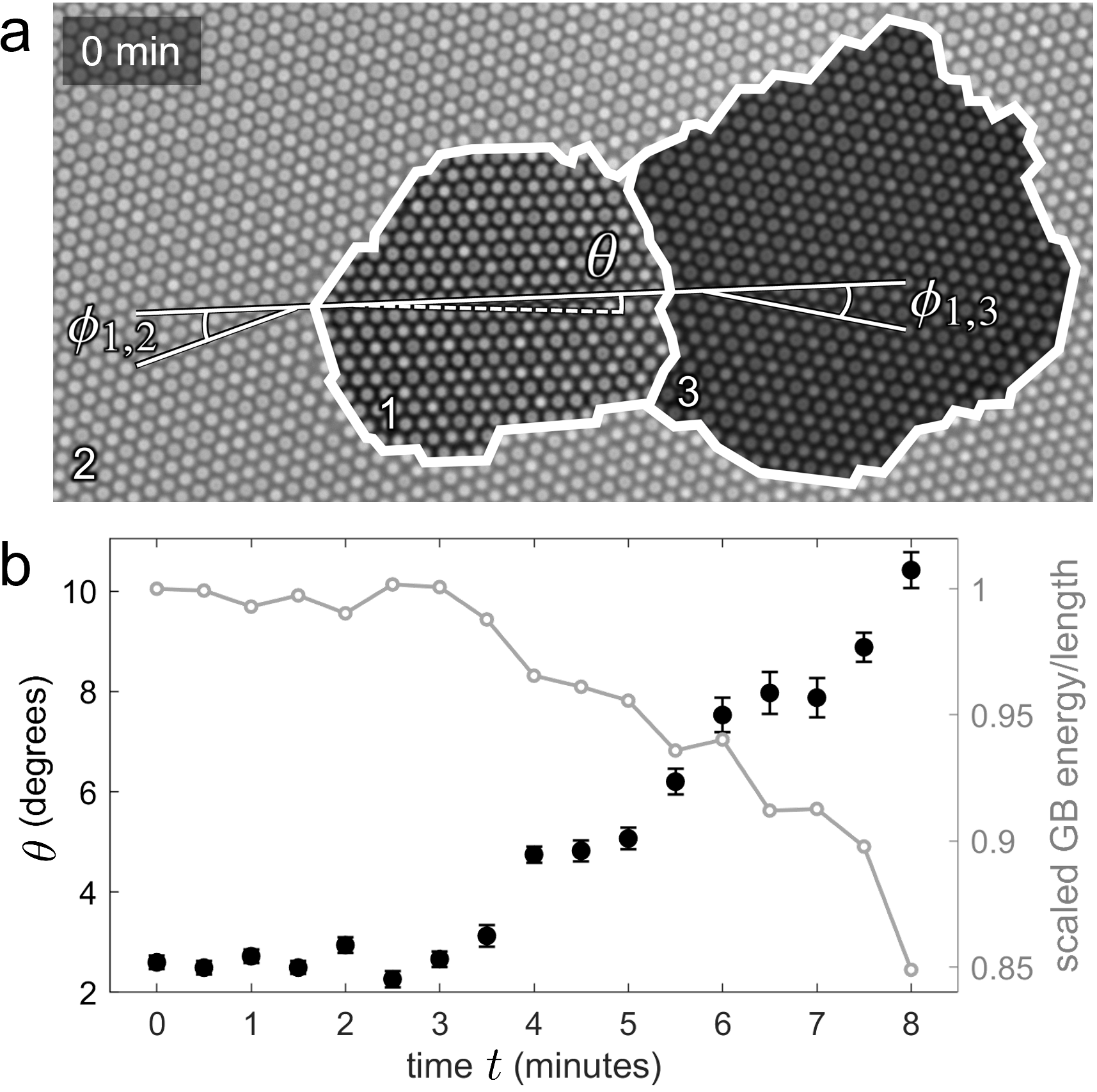}
\caption{A grain rotates to better match the orientation of its larger neighbor in a simple 2D colloidal polycrystal. (a) Grain 1 (center of the microscope image) is oriented at an angle $\theta$ relative to the horizontal axis (dotted line), and is adjacent to grain 2 (shaded dark) and grain 3 (shaded light). The polycrystal is shown here at $t = 0$, and grain boundaries are highlighted with bold white outlines. The misorientation angles between grain 1 and its neighbors are $\phi_{1,2}$ and $\phi_{1,3}$, respectively. (b) Over time, grain 1 rotates counterclockwise to better match the orientation of grain 2. Grain 1's orientation $\theta$ is plotted with black markers. Error bars indicate standard error of the mean (SEM) obtained by averaging over all particles in the grain. The rotation reduces the Read-Shockley energy per unit length (gray line and open circles, scaled GB energy) relative to its $t=0$ value.}
\label{fig:Rotation}
\end{figure}

As a first step toward studying polycrystalline systems in which each crystal grain moves under the possibly competing influences of multiple neighbor grains, we use optical blasting to assemble two adjacent grains (labelled 1 and 3 in Fig.~\ref{fig:Rotation}a), both embedded within a larger crystal grain (labelled 2 in Fig.~\ref{fig:Rotation}a). We use this simple polycrystalline grain configuration to explore how grain 1 moves in response to interactions with its two neighbors: grain 2, whose lattice is oriented counterclockwise relative to grain 1; and grain 3, whose lattice is oriented clockwise relative to grain 1. This grain configuration is observed at a rate of 2 frames/minute.

As described in the sections below, we find that in response to the competing interactions with its two neighboring grains, grain 1 first steadily rotates as a rigid object, and then abruptly splits apart into two counterrotating regions.

\section{\label{sec:Rotation}Grain rotation}
We observe evidence of grain rotation within our colloidal polycrystalline grain configuration. Over the course of about 8 minutes, grain 1 in Fig.~\ref{fig:Rotation}a shrinks and also rotates counterclockwise as a rigid object to better match the orientation of grain 2. This rotation lowers the free energy; we estimate the reduction using the Read-Shockley grain boundary energy equation, a continuum description which, in two dimensions, defines the cost per unit length of a grain boundary segment as $\gamma(\phi) = \gamma_0 \phi (A - \ln\phi)$, where $\phi$ is the misorientation angle and $A$ and $\gamma_0$ are constants determined by the elastic moduli of the crystal \cite{Read1950}. In our colloidal system, $A \approx 1$ \cite{Cash2018}. As grain 1 rotates counterclockwise, it becomes more aligned with grain 2 and more misaligned with grain 3. The Read-Shockley energy per unit length decreases overall (Fig.~\ref{fig:Rotation}b) because the boundary with grain 2 is longer than the boundary with grain 3. Indeed, an effective torque on grain 1 due to each of its neighbors can be computed from the derivative of the Read-Shockley energy with respect to the misorientation angle: $\tau_{\mathrm{eff}} = s~ {\mathrm{d}\gamma}/{\mathrm{d}\phi} $ is the torque on a length $s$ of a grain boundary with misorientation angle $\phi$ \cite{Harris1998,Moldovan2001}. Computing the effective torques applied by the neighboring grains, we find that throughout grain 1's rotation, the counterclockwise torque $\tau_{1,2}$ exerted by grain 2 is always more than twice the clockwise torque $\tau_{1,3}$ exerted by grain 3. This torque imbalance causes the observed counterclockwise rotation, in accordance with the Read-Shockley theory.

\begin{figure}[h!] \includegraphics[width=0.48\textwidth]{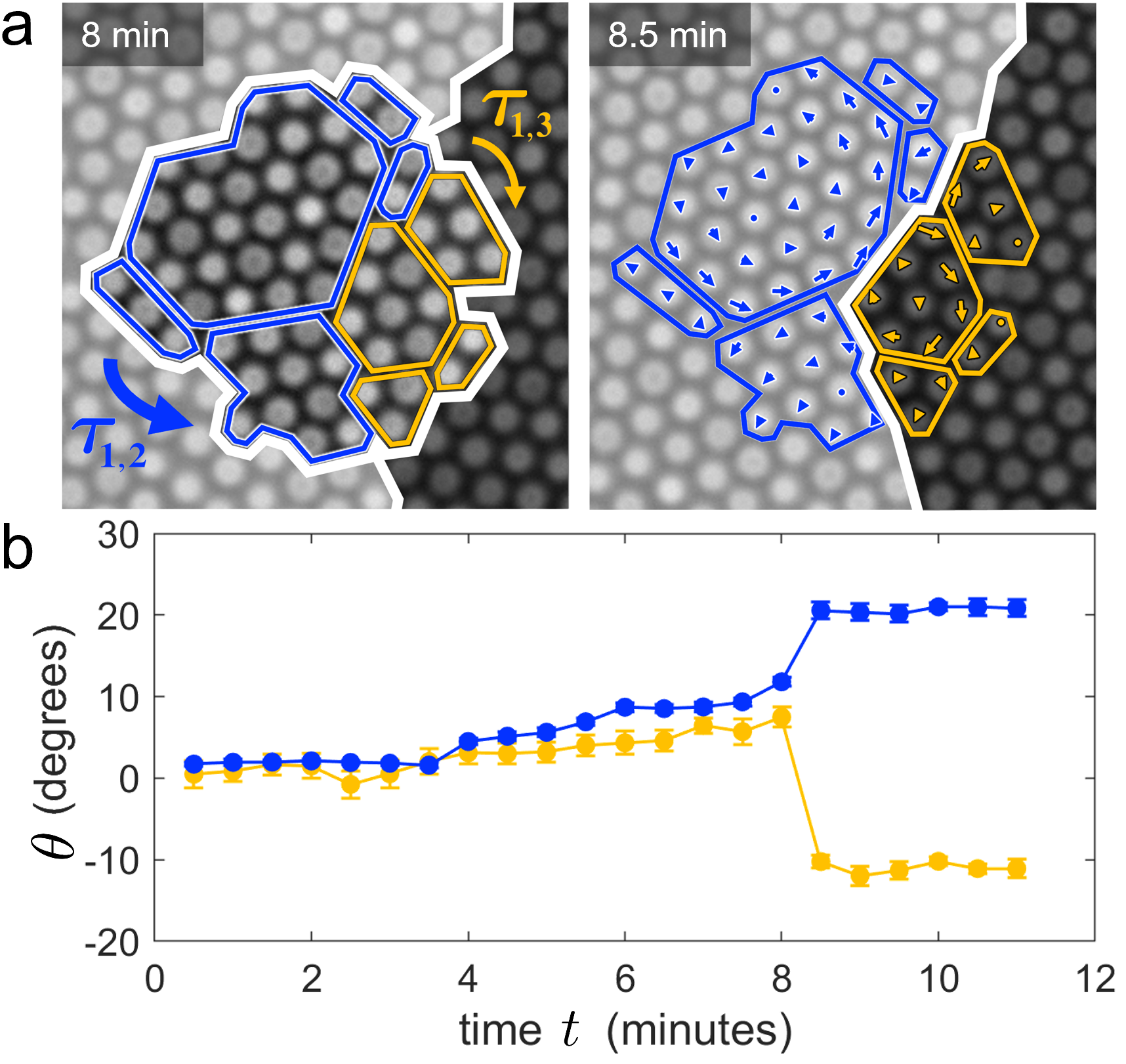}
\caption{A colloidal crystal grain splits to coalesce with its neighboring grains. (a) A central grain (grain 1) experiences opposing torques (curved arrows) and splits into two counterrotating regions. During splitting, individual granules outlined in blue and yellow rotate independently (arrows indicate particle displacements; circles indicate minimal displacements smaller than the size of an arrowhead). (b) The local crystal orientations around the particles from the two counterrotating regions (counterclockwise in blue, clockwise in yellow) abruptly change during splitting. Both regions initially rotate counterclockwise as a rigid object, and then rapidly rotate in opposite directions. This rapid rotation splits the grain, with each of the two regions coalescing with its adjacent grain. Error bars indicate SEM obtained by averaging over orientations of individual granules. }
\label{fig:Splitting}
\end{figure}

\section{\label{sec:Splitting}Grain splitting}

While the observed grain rotation is well described by conventional grain torque theories, we also observe a microscopic mechanism of grain coarsening that is beyond their purview. Fig.~\ref{fig:Splitting} shows grain 1 from Fig.~\ref{fig:Rotation}, after shrinking and rotating over the course of 8 minutes. Grain 2 and grain 3 still exert opposing torques on grain 1, but in this case grain 1 does not rotate as a rigid object. Instead, as shown in Fig.~\ref{fig:Splitting}, the grain splits apart into two counterrotating regions. The left region (blue outlines) rotates counterclockwise to match the orientation of grain 2, while the right region (yellow outlines) rotates clockwise to match the orientation of grain 3. We observe that this splitting reduces the free energy of the system by 60 percent, as approximated by the Read-Shockley grain boundary energy.

\subsection{\label{sec:granules}Granule formation}
Remarkably, the counterrotation of these two regions occurs via the independent rotation of individual granules, each composed of very few particles rotating together as a rigid object. Fig.~\ref{fig:Splitting}a outlines each granule and indicates the displacements of individual particles with arrows. To track the rotation of the two counterrotating regions, we measure the average orientation of the counterclockwise/clockwise (blue/yellow) splitting particles throughout the experiment. Both sets of particles slowly rotate counterclockwise as described in the preceding section, until $t=8$~min, when the grain abruptly splits.

\begin{figure}[h!] \includegraphics[width=0.48\textwidth]{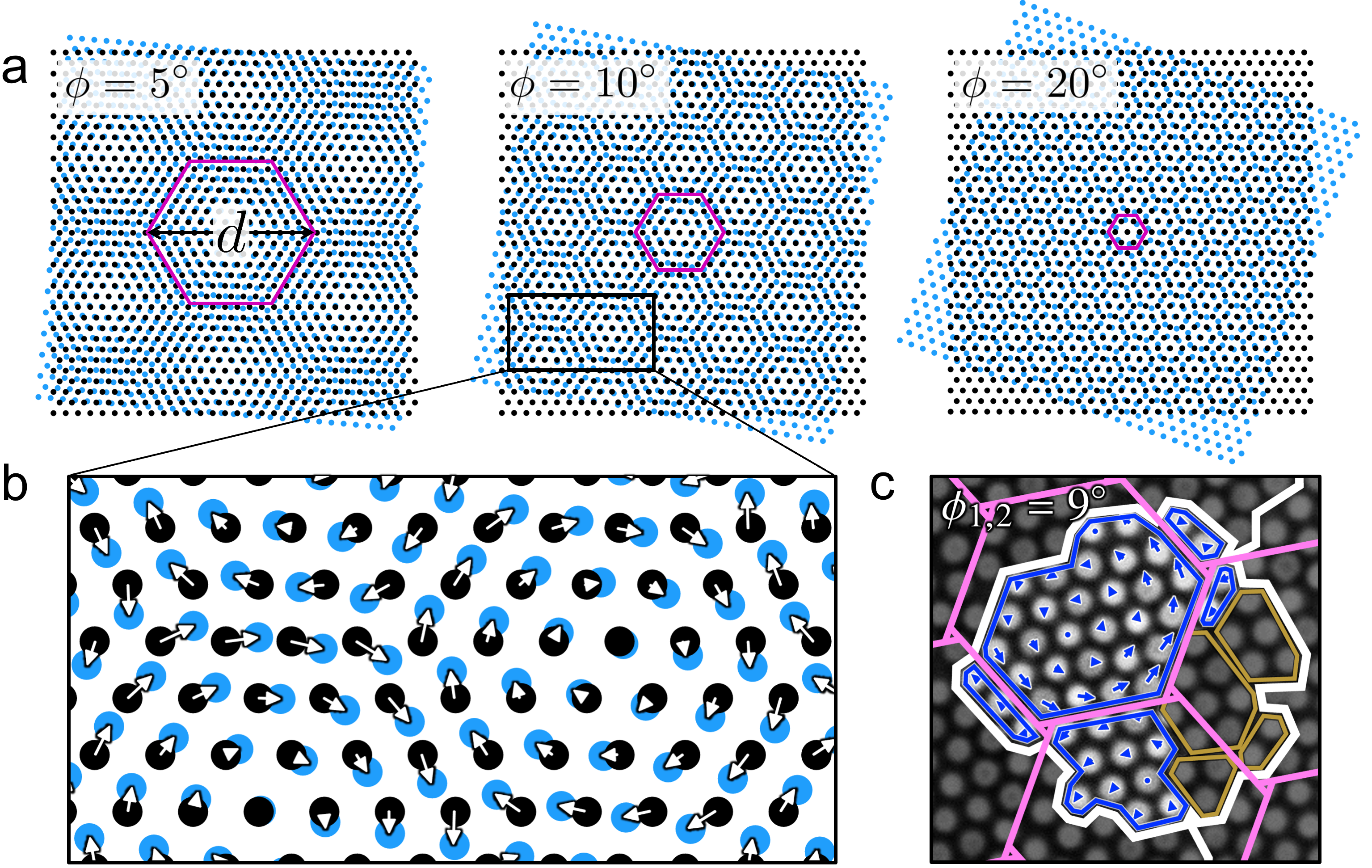}
\caption{Rotating granules are determined by the underlying Moir\'e pattern. (a) Two overlaid triangular lattices with misorientation angle $\phi$ form a Moir\'e pattern with hexagonal regions. For each pattern, an ideal granule is outlined. The diameter $d$ of an ideal granule decreases with increasing $\phi$. (b) When each black lattice site is matched to the nearest blue lattice site, the displacements show clockwise rotation about the center of each ideal granule. (c) In the experiment from Fig.~\ref{fig:Splitting} at $t=8$~min, the lattices of grains 1 and 2, which have a misorientation angle of $\phi_{1,2}=9^{\circ}$, determine ideal granules (outlined in pink). The counterclockwise-rotating granules are fragments of ideal granules that have been cut off by neighboring boundaries.}
\label{fig:Moire}
\end{figure}

These individually rotating granules form because grain splitting takes place over a much shorter time scale than rigid-object grain rotation. Consider a grain rotating as a rigid object. Particles farther from the center of rotation must move a greater distance, which limits rotation speed. However, the same change in crystal orientation can also be achieved with much shorter displacements. This is accomplished when particles move to the closest final lattice position, so that multiple hexagonal regions form, each rotating about its own center. This is illustrated in Fig.~\ref{fig:Moire}a-b, where an initial lattice (black points) is overlaid with a rotated lattice (blue points), creating a Moir\'e pattern with hexagonal regions (pink). The diameter $d$ of these hexagonal ``ideal granules'', measured in lattice constants (LC), is determined geometrically and varies as a function of the misorientation angle $\phi$ between the lattices as $d=\left(1+\cos\phi\right)/\left(\sqrt{3}\sin\phi\right)$, with lower misorientation angles corresponding to larger hexagons. Note that for $d<3$, granules can no longer be clearly defined.

The granules observed in the grain splitting experiment (Fig.~\ref{fig:Splitting}) can be understood by considering the ideal granules set by the underlying Moir\'e patterns. This is shown in Fig.~\ref{fig:Moire}c, where the $t=8\,$min experimental image is overlaid with the ideal granules (outlined in pink) determined by the crystal lattices of grains 1 and 2. The misorientation between these lattices is $9 ^\circ$, corresponding to ideal granules with diameter $d=7.3~$LC. Comparing the pattern of ideal granules with the experimentally observed granules, we see that the experimental counterclockwise-rotating granules (blue) are exactly bounded by the ideal granule boundaries, the grain boundary (white), and the boundaries of the clockwise-rotating granules (yellow). Although the experimental granules are not perfect hexagons, the particle trajectories during splitting are determined by the underlying Moir\'e patterns. A similar mechanism for grain growth has been previously proposed  \cite{Srinivasan2002,Cahn2004}, but never directly observed. 

\subsection{\label{sec:EnergyBarrier}Free energy barrier to grain splitting}

Although grain splitting ultimately reduces the free energy, the system must first overcome an energy barrier. During grain splitting, many tiny grain boundaries form between the granules. To estimate the free energetic cost of creating these granule boundaries, we cannot use the Read-Shockley equation, because the granules each contain few particles and therefore are not well described by a continuum theory. Instead we introduce a model for the Helmholtz free energy of the hard sphere colloidal crystal.

\subsubsection{Model for the free energy of a colloidal crystal grain}
\label{sec:theory}

\begin{figure}[h!] \includegraphics[width=0.48\textwidth]{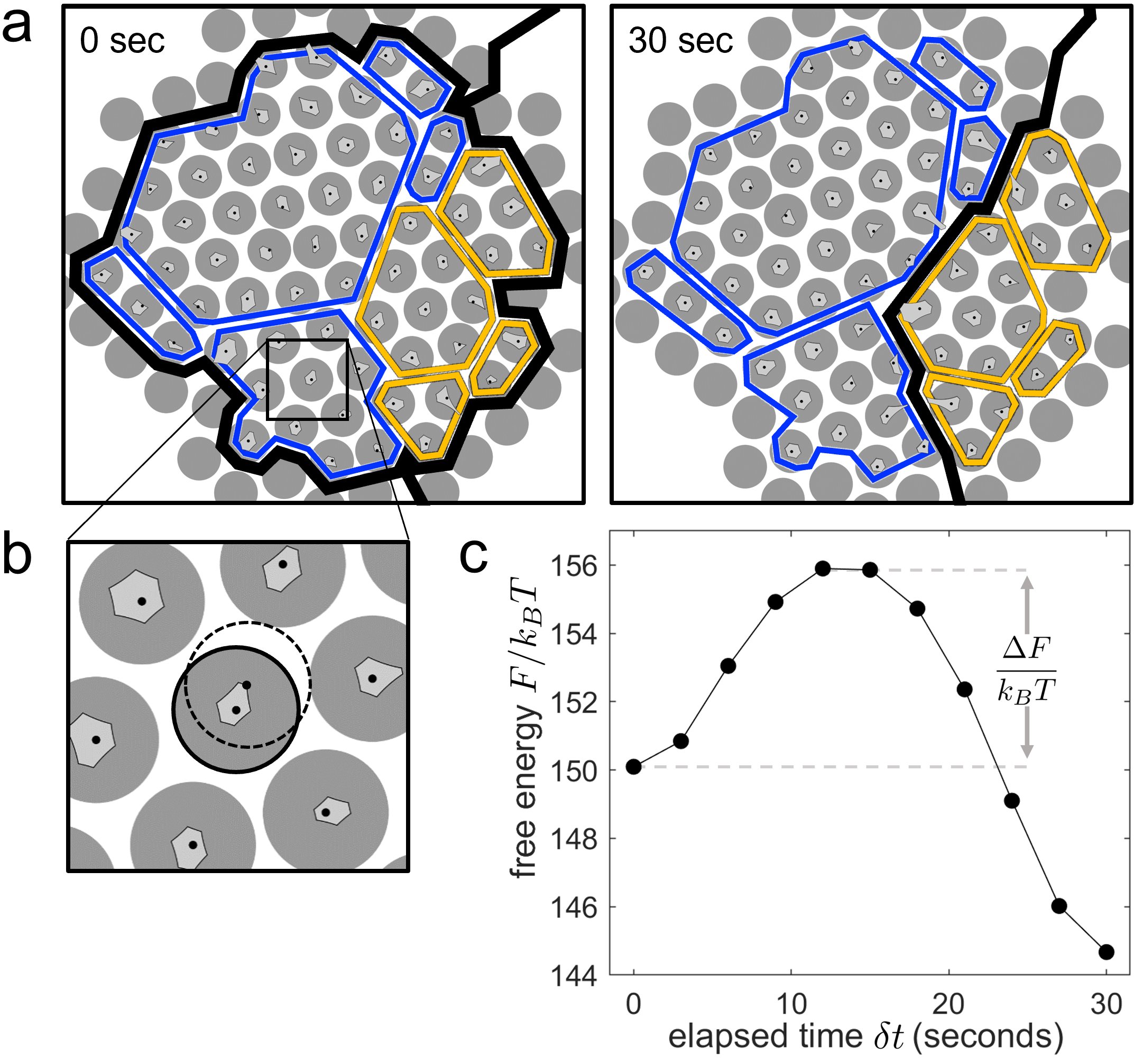}
\caption{The Helmholtz free energy of the grain increases before decreasing during the experimentally observed grain splitting.  (a) Each particle's free space (light gray shape at the center of each particle) is shown before and after grain 1 splits. (b) A single particle is highlighted, showing that the edge of the particle's free space corresponds to the farthest position the particle could move to (dotted particle outline) without overlapping its neighbors in their current positions. (c) Helmholtz free energy over time as the grain splits, assuming that particles move along straight trajectories. Grain splitting lowers the energetic cost of the grain boundaries, but first requires an increase in energy. Here the elapsed time $\delta t = t - 8$~min.}
\label{fig:FreeEnergyExp}
\end{figure}

We calculate the Helmholtz free energy $F=U-TS$, where $U=0$ for the hard sphere colloidal particles, $T$ is the temperature, and $S$ is the entropy. We model the 2D colloidal polycrystal as a collection of hard disks of radius $R$, and compute its entropy as $S=k_B\sum_i \ln\frac{v_i}{\pi R^2}$, where we sum over every particle in grain 1. Here, $k_B$ is the Boltzmann constant and $v_i$ is the area of free space available to the $i^\text{th}$ particle, that is, the space that can be accessed by the particle’s center without the particle overlapping with any other particles in their current positions (an example is shown in Fig.~\ref{fig:FreeEnergyExp}b). This formulation of entropy correctly predicts that, in a dense suspension of colloidal hard spheres, the most statistically favorable way to arrange the particles is on a triangular lattice where the particles have the most free space on average. Similar approaches have been previously applied to counting microstates of hard sphere systems \cite{speedy, LJDbook, weeks3disks, weeks4disks}.

\subsubsection{Determining the energy barrier}
Though we have the experimental particle positions from before and after the splitting event, we must interpolate the particle positions between those two points in time to find the free energy barrier. We assume each particle moves in a straight line at a constant rate from their pre-splitting positions to their post-splitting positions. At each timestep, we calculate the free energy as $F=-k_BT\sum_i \ln\frac{v_i}{\pi R^2}$. As shown in Fig.~\ref{fig:FreeEnergyExp}, this estimated free energy initially increases as the granules rotate, since granule rotation introduces many new granule boundaries. Then the free energy ultimately decreases as the granules align into their final crystal orientations. Overall, during the experimentally observed splitting event, the system overcomes an energy barrier of height $\Delta F=5.8~k_BT$, indicating that this barrier is likely to be overcome by a typical thermal fluctuation.

\subsubsection{Effect of grain size and misorientation angle}
Grain coarsening via grain splitting has not been previously reported, to our knowledge. To investigate the range of situations in which grain splitting may occur, we studied how the height of the free energy barrier depends on both the diameter $D$ of the central grain and the misorientation angle $\phi$ between that grain and its two neighbors. We simulated grain splitting events in which a circular central grain is neighbored by a left grain and a right grain with equal and opposite misorientation angles, as shown in Fig.~\ref{fig:SimSplitting}. Complete simulation methods are provided in the Supplemental Material \cite{suppmat}. Briefly, we find the lattice positions associated with each grain, and we translate the lattices of the right grain and central grain to maximize the number of non-overlapping particles that fit in the simulation window. Then we match the particles from the central grain to final positions in the left or right grain by sequentially matching each particle with the nearest unoccupied lattice site. When matching, we prioritize particles with a bigger difference in the distance between nearest unoccupied lattice site and second-nearest unoccupied lattice site. The result is that the central grain splits into granules that rotate according to the underlying Moir\'e pattern. For sufficiently small misorientation angles and grain diameters, there is not enough room within the central grain for multiple granules to form, so we exclude this very low $D$, low $\phi$ (high $d$) region of the phase space ($d \gtrsim 4D/5$). For each misorientation angle $\phi$ and grain diameter $D$, we compute the free energy at 11 timesteps. We find that the simulated system always overcomes a free energy barrier during splitting.
 
\begin{figure}[h!] \includegraphics[width=0.48\textwidth]{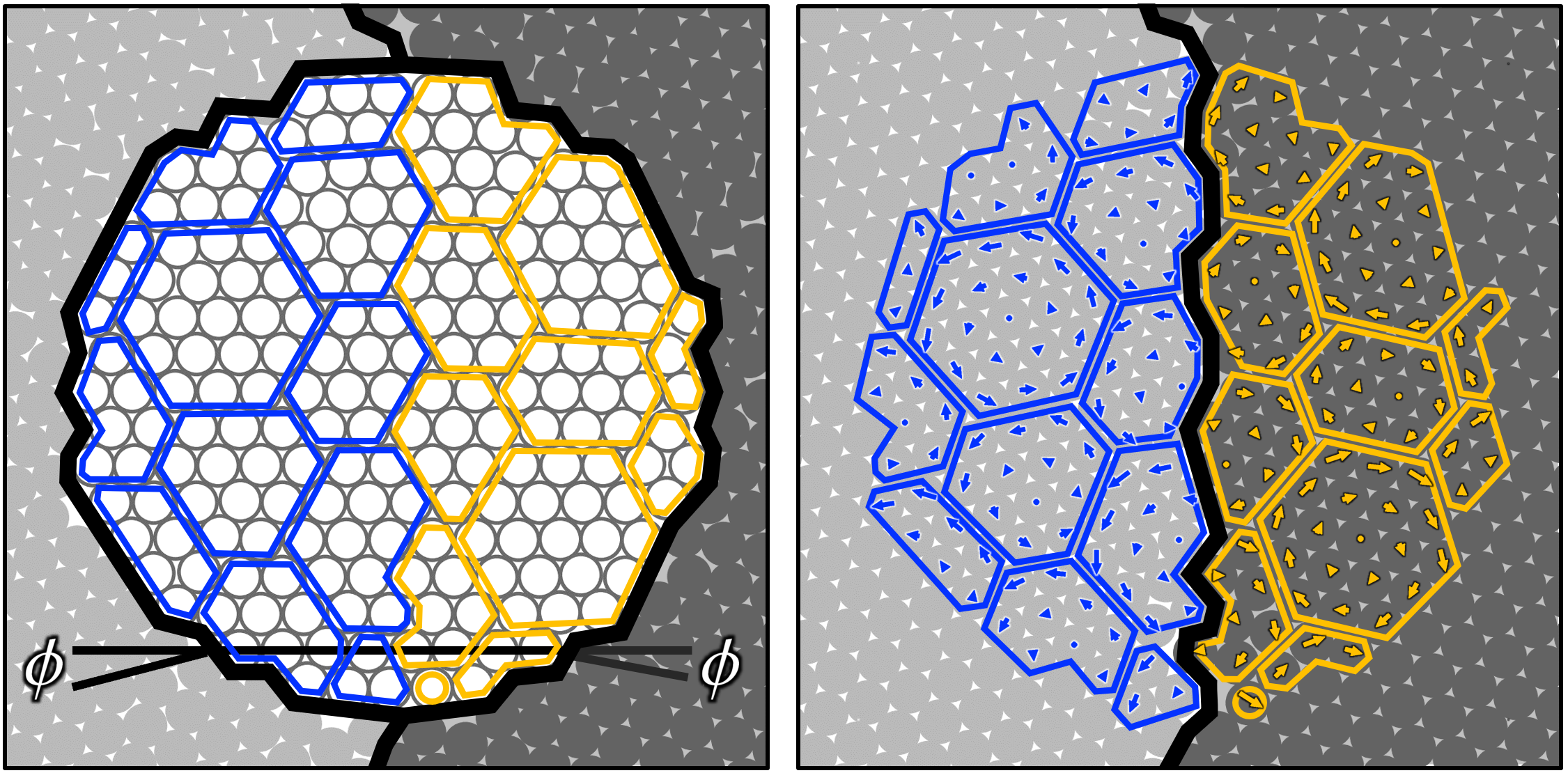}
\caption{Grain splitting events were simulated to determine how the free energy barrier depends on grain diameter $D$ and misorientation angle $\phi$. An example simulated grain splitting event with $D=15~$LC and $\phi=15^\circ$ is shown. A central grain (white particles) is initially neighbored by grains (light and dark gray particles) with equal and opposite misorientation angles $\phi$. The central grain splits via granules that rotate counterclockwise (blue) and clockwise (yellow) according to the underlying Moir\'e pattern. Particle displacements are shown with blue and yellow arrows.}
\label{fig:SimSplitting}
\end{figure}

\begin{figure}[t!] \includegraphics[width=0.47\textwidth]{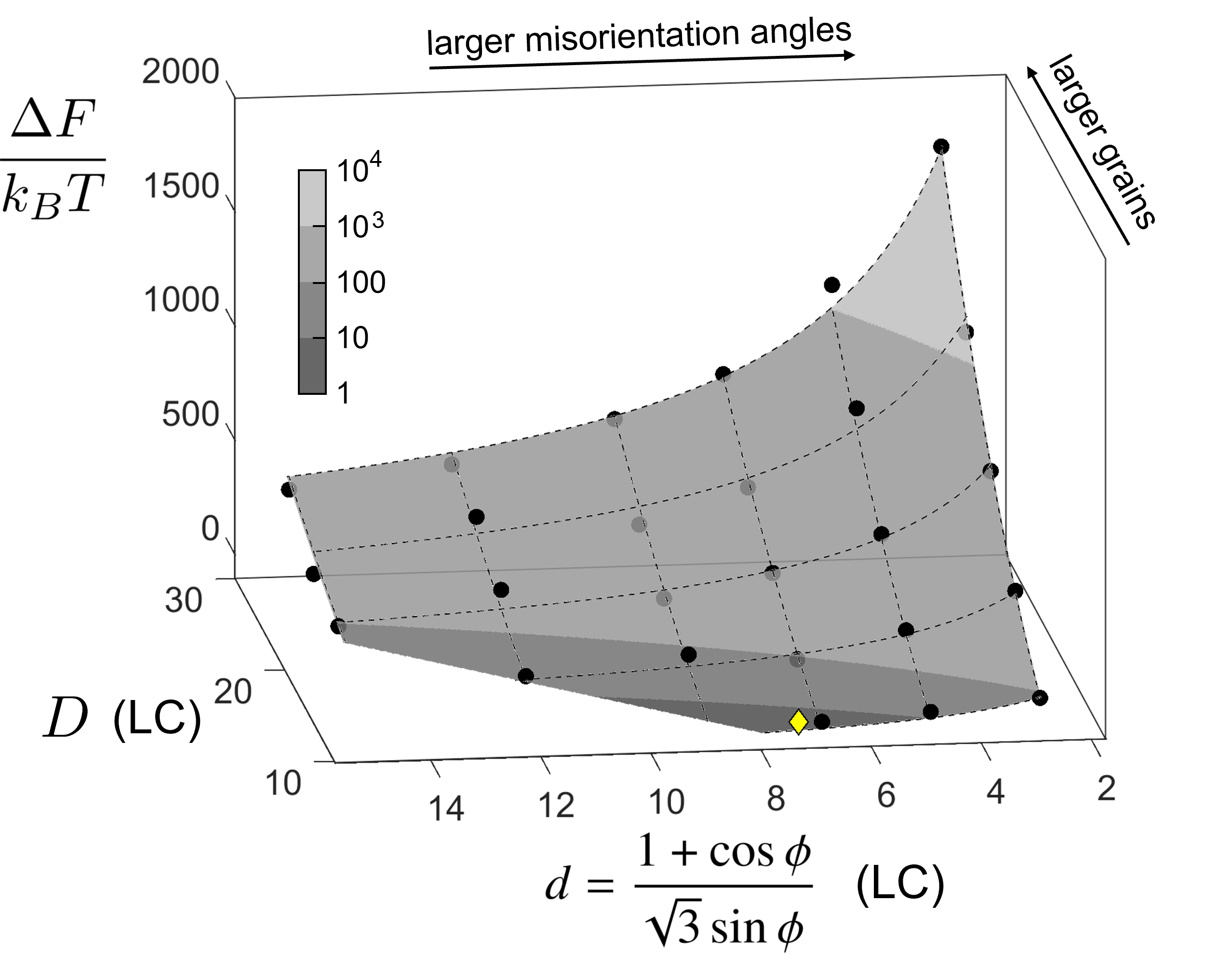}
\caption{
The barrier to grain splitting increases with increasing grain diameter $D$ and decreasing ideal granule diameter $d$ (increasing misorientation angle $\phi$). The shaded surface plot with dashed lines represents the prediction that the energy barrier scales like  $L \approx \sqrt{3}D\left(D/d + 1\right)$, an approximation for the total length of granule boundaries. The black data points are the free energy barrier heights $\Delta F / k_B T$ as directly computed from grain splitting simulations. The single yellow diamond data point represents the energy barrier from the experimentally observed splitting event (Fig.~\ref{fig:FreeEnergyExp}c). Shades of gray correspond to values of $\Delta F / k_B T$, as indicated by the colorbar.}
\label{fig:BarrierScaling}
\end{figure}

As shown in Fig.~\ref{fig:BarrierScaling}, the free energy barrier to grain splitting $\Delta F / k_B T$ increases with both the diameter of the grain $D$ and the misorientation angle $\phi$ between the central grain and the two neighboring grains. These trends can be understood intuitively by considering the total length of granule boundaries formed during the splitting event. Higher misorientation angles result in smaller granules, and thus more granule boundaries for a set grain diameter $D$. For a fixed misorientation angle, larger grains divide into more granules, again leading to more granule boundaries. We can estimate how the total length of granule boundaries $L$ depends on $\phi$ and $D$ by approximating that the grain divides into whole hexagonal ideal granules. Then $L \approx \sqrt{3}\left(\frac{D^2}{d} + D\right)$, where $d =\left(1 + \cos \phi\right)/\left(\sqrt{3}\sin\phi\right)$ is the ideal granule diameter determined by $\phi$. The free energy barrier $\Delta F / k_B T$ should scale like $L$. This scaling prediction is plotted as a surface in Fig.~\ref{fig:BarrierScaling}, linearly fit to the black data points determined by directly calculating $\Delta F / k_B T$ from the grain splitting simulations.

Only a small range of $d$ and $D$ values allows for a grain splitting barrier that is accessible within a few $k_BT$, so that a typical thermal fluctuation could overcome the barrier. For example, a grain splitting event with $d$ and $D$ values falling outside of the darkest shaded region in Fig.~\ref{fig:BarrierScaling} is at least 60 times more unlikely than the event we experimentally observed (yellow diamond in Fig.~\ref{fig:BarrierScaling}). For $d$ and $D$ values outside the second darkest region, events are over $10^{41}$ times more unlikely. As such, we expect grain splitting events to be rare, except in systems with very small grains with relatively low misorientation angles.

\section{Conclusion}
We have performed the first experimental study of a 2D colloidal polycrystal in which a single crystal grain experiences competing torques caused by its two neighbor grains. We have observed that this grain rotates in the direction of the net torque, as predicted by established continuum theories. However, this conventional framework cannot explain our observation of grain splitting, in which the grain separates into two counterrotating regions composed of multiple independently rotating granules. Furthermore, because the granules contain so few particles, the continuum Read-Shockley model for the free energy of a grain boundary is insufficient to describe the free energy barrier to grain coarsening via grain splitting. We have directly computed the entropy and free energy of the system during splitting, finding that there is a barrier that increases with grain size and grain misorientation. While we determined this free energy barrier for our hard sphere system, it persists upon adding an attractive interatomic potential, such as Lennard-Jones, because atomic bonds stretch as granules rotate \cite{suppmat}. Grain splitting is likely to be a significant mechanism for grain coarsening in the regime of small grains and low misorientation angles. Consequently, grain growth in nanocrystalline materials may be better described by incorporating the mechanism of grain splitting into mesoscale models in this regime.

\acknowledgements{The authors thank J. M. Gregoire, N. P. Breznay, and A. L. Cook for helpful comments on the manuscript, and J. Abacousnac and R. L. Barcklay for contributions in the lab. This work was funded by the Research Corporation for Science Advancement through a Cottrell Scholar award to SJG, and computations were supported by XSEDE supercomputing resources. We also acknowledge Harvey Mudd College for internal funding through the J. Arthur Campbell Summer Research Fund and the Harvey Mudd College Physics Summer Research Fund. }

\bibliography{GrainsSplit}
\end{document}


\setcounter{figure}{0}
\renewcommand{\thefigure}{S\arabic{figure}}

\preprint{APS/123-QED}

\title{Supplemental materials \\
Grain splitting is a mechanism for grain coarsening in colloidal polycrystals}

\author{Anna R. Barth}
\thanks{These authors contributed equally.}
 \affiliation{%
 Department of Physics, Harvey Mudd College, Claremont, CA
}%

\author{Maya H. Martinez}
\thanks{These authors contributed equally.}
 \affiliation{%
 Department of Physics, Harvey Mudd College, Claremont, CA
}%

\author{Cora E. Payne}
 \affiliation{%
 Department of Physics, Harvey Mudd College, Claremont, CA
}%

\author{Chris G. Couto}
 \affiliation{%
 Department of Physics, Harvey Mudd College, Claremont, CA
}%

\author{Izabela J. Quintas}
 \affiliation{%
 Department of Physics, Harvey Mudd College, Claremont, CA
}%

\author{Inq Soncharoen}
 \affiliation{%
 Department of Physics, Harvey Mudd College, Claremont, CA
}%

\author{Nina M. Brown}
 \affiliation{%
 Department of Physics, Harvey Mudd College, Claremont, CA
}%

\author{Eli J. Weissler}
 \affiliation{%
 Department of Physics, Harvey Mudd College, Claremont, CA
}%

\author{Sharon J. Gerbode}
 \email{gerbode@hmc.edu}
\affiliation{%
 Department of Physics, Harvey Mudd College, Claremont, CA
}%

\maketitle

\section{Simulating grain splitting events}

Our grain splitting simulations consist of a circular central grain neighbored on either side by two other grains with equal and opposite misorientation angles. The central grain splits in half vertically to match the orientations of its neighbors, as shown in Fig.~5 of the main paper. To set up these simulated grain splitting events, we first create the left and right grains. We start by defining the left crystal lattice in the left half-plane, and then select the position for the right crystal lattice that maximizes the number of particles that fit without overlapping by more than half a particle radius. These small overlaps are later resolved as described below. Then, we define a circular boundary of diameter $D$ and within this boundary, we center a third lattice corresponding to the central grain. The central grain boundary is centered at the midpoint between the two neighbor grains, and is vertically translated to minimize the area of overlapping particles and avoid producing vacancies in the final configuration. We fill the third central lattice with particles at each lattice site, and seek to find final lattice positions for those particles.

\begin{figure}[ht]
 \includegraphics[width=0.48\textwidth]{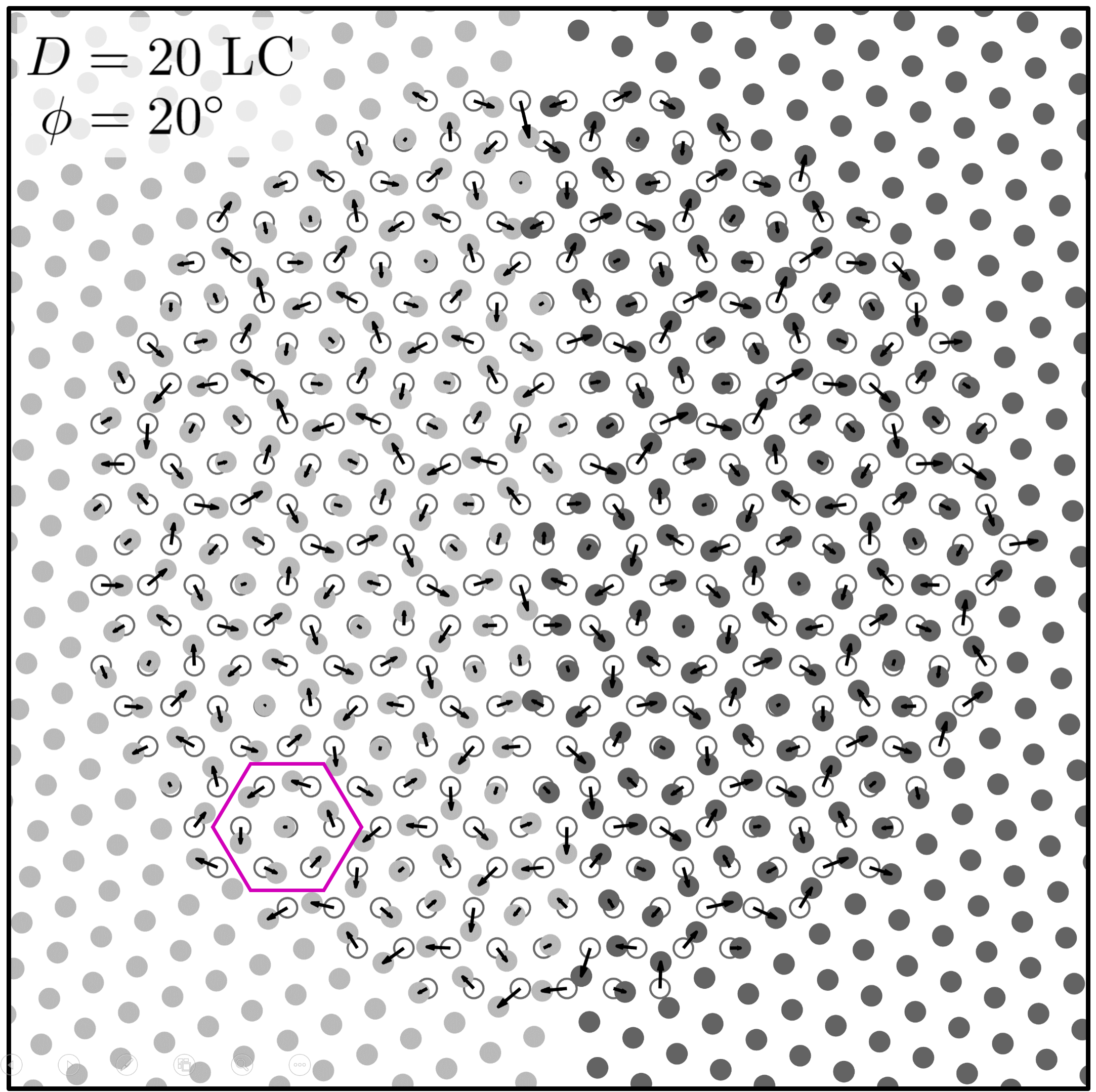}
  \caption{Particles from the central grain (white circles with gray outlines) are matched to new lattice positions (light and dark gray circles) according to the underlying Moir\'e patterns. Here, the central grain has diameter $D=20~$lattice constants, and the misorientation angle between the central grain and each of its neighbors is $\phi=20^\circ$. An ideal granule has been outlined in pink.
  }
  \label{fig:Matching}
\end{figure}

We match the particles in the central grain to new lattice positions corresponding to the left or right grain, forming rotating granules according to the underlying Moir\'e patterns. In brief, this is accomplished by matching the particles to nearby lattice sites. This matching prioritizes particle placement by considering the fractional difference between the closest unoccupied final lattice site and the second closest unoccupied final lattice site. Particles with large fractional differences are given priority when matching, so that the particles that are much closer to one possible site than another will be placed in the closest site, while particles that are nearly equidistant to two available lattice sites are placed later. Further considerations include minimizing particle overlap and avoiding vacancies in the final particle configuration. In the event particle trajectories are found to cross, the initial vertical position of the central grain is adjusted to prevent this, and then particles are again matched to new final lattice positions. The result of this matching algorithm is depicted in Fig.~\ref{fig:Matching} for $D=20$~LC and $\phi=20^\circ$.

After matching, the initial and final configurations are checked for overlapping particles, which are corrected by moving each particle back equally along the line connecting their centers until there is no overlap. Then, we construct 9 intermediate timesteps by advancing particle positions along straight lines from the initial to final configurations at a constant rate. At each timestep, any overlapping particles are resolved and the free energy $F = U - TS$, is computed, with $S = k_B\sum_i \ln \frac{v_i}{\pi R^2}$, where we sum over the particles in the central grain. Here, $v_i$ is the free space available to the $i^\mathrm{th}$ particle and $R$ is the particle radius, as described in the main paper, in Section~IVB.

\section{Finding grain boundaries using the minimum cut algorithm}

\begin{figure}[ht]
 \includegraphics[width=0.48\textwidth]{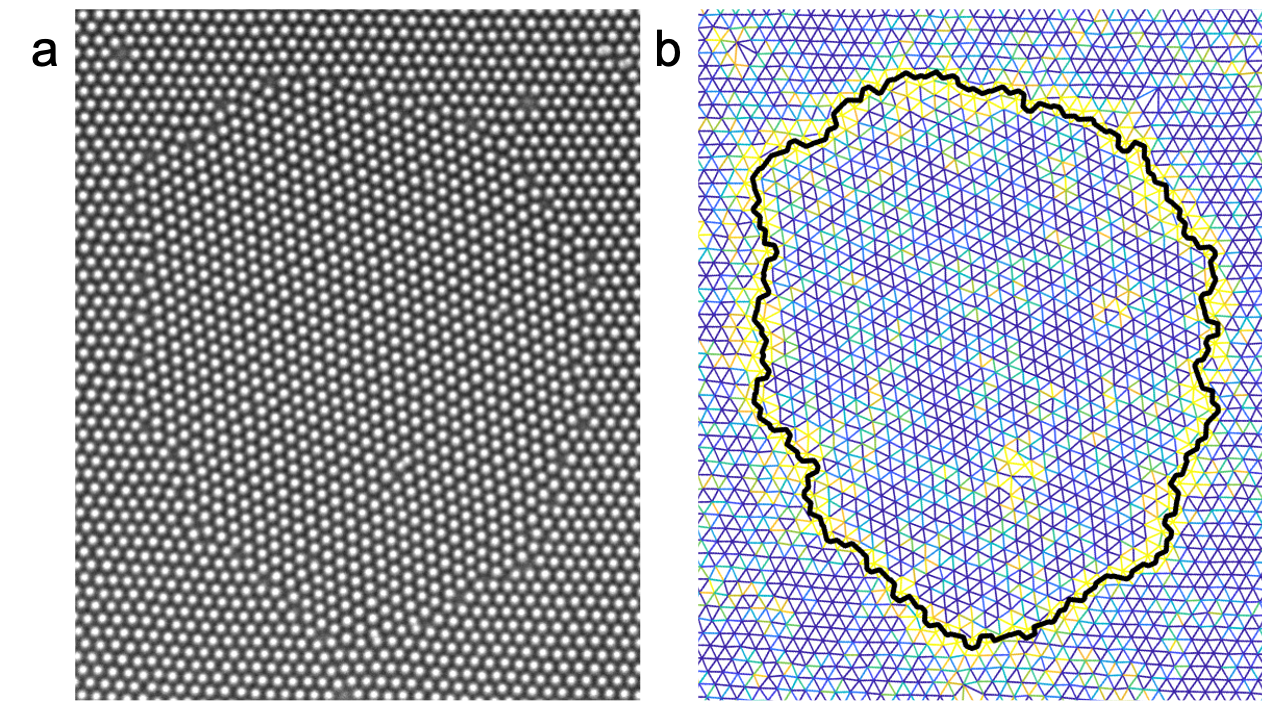}
  \caption{Grain boundaries are located using the minimum cut algorithm. (a) A grain boundary loop encloses a colloidal crystal grain surrounded by a neighbor grain with a different orientation. (b) Weighted graph representation and calculated grain boundary (black line). Vertices correspond to particle centers and edges are colored according to their weights using a logarithmic colormap.
  Edges near the grain boundary have weights an order of magnitude less than those elsewhere in the graph. 
  }
  \label{fig:GBfind}
\end{figure}

To locate grain boundaries computationally from microscope images, we represent the colloidal crystal as a weighted, undirected graph. Each colloidal particle is represented as a vertex in the graph, connected to its nearest neighbors by edges that are weighted by the similarity of the local crystal orientation of connected particles. For each particle, we quantify local crystal orientation with the phase angle $\Theta$ of the $\Psi_6$ local orientational parameter, defined in terms of a sum over $N$ nearest neighbors as $\Psi_6=\frac{1}{N}\sum_{n=1}^Ne^{i6\theta_n}\equiv |\Psi_6|e^{i\Theta}
$, where $\theta_n$ is the angle of the vector between one particle and its $n^\mathrm{th}$ nearest neighbor, measured from the horizontal axis. The weight of the edge between neighboring particles $i$ and $j$ is computed as $
    w_{ij} 
    = 
    |\Theta_i - \Theta_j|^{-k}$,
where $\Theta_i$ and $\Theta_j$ are the phase values of the $\Psi_6$ local orientational parameter computed for particles $i$ and $j$, respectively, and $k = 20$ is a positive exponent whose exact value does not significantly influence the grain boundary obtained. Using this weight tensor, edges between particles in the same grain will have large weights, while edges between particles in different grains will have small weights. Fig.~\ref{fig:GBfind}a depicts a single colloidal grain embedded within another crystal, while Fig.~\ref{fig:GBfind}b shows the corresponding weighted graph; the edges are colored according to the logarithm of their weights and each vertex corresponds to the center of a colloidal particle.

We then partition the graph into two pieces using a minimum cut algorithm, which is often used in image segmentation \cite{Boykov2001}. To partition a graph between vertices $i$ and $j$,  edges are removed from the graph until there exists no path between those two vertices along the graph edges. The minimum cut of a graph is a partition between two vertices in which the sum of the weights of the removed edges is minimized. The minimum cut of a graph of our colloidal crystal, then, removes the low-weight edges that lie between particles in different grains, thus separating the graph into pieces corresponding to different grains. By choosing one vertex (particle) to be inside the grain boundary loop and the other to be outside, we use the minimum cut algorithm to isolate the inner grain from the outer grain, allowing us to identify the shared Voronoi polygon edges that compose the grain boundary (black line in Fig.~\ref{fig:GBfind}b). To prevent erroneous partitioning immediately surrounding the ``seed'' vertices, we first inflate the weights of the edges immediately connected to those vertices. The minimum cut method of grain boundary location is 200 times faster than our implementations of existing methods \cite{Cash2018, Skinner2010}, while not sacrificing accuracy. Unlike those methods, the minimum cut algorithm can process grain boundaries of arbitrary shape, as the crystal is represented only using nearest neighbor connections.

\section{Adding an interparticle potential}
To test how grain splitting might work in a non-hard-sphere system, where the Helmholtz free energy has a nonzero internal energy term $U$, we explored how this internal energy would evolve during a grain splitting event, assuming a Lennard-Jones interaction between the particles. At each timestep of the same grain splitting simulations in the main paper, we computed $U = \epsilon\left( (r_0/r)^{12}-2(r_0/r)^6 \right)$ where $\epsilon$ is the energy scale of the interaction, and $r_0$ is the energy-minimizing separation, which was taken to be $1\,$LC. 

\begin{figure}[ht]
 \includegraphics[width=0.48\textwidth]{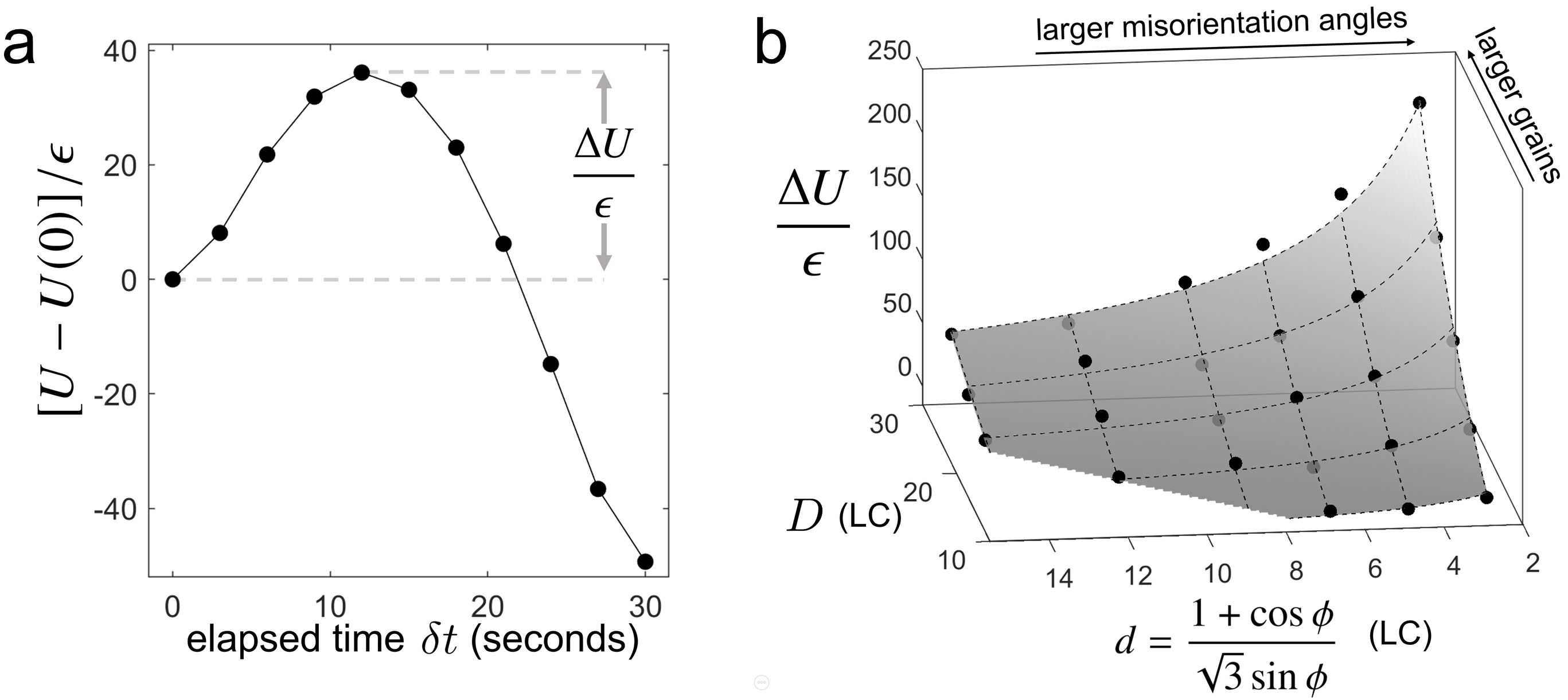}
  \caption{Grain splitting requires the system to overcome a barrier in Lennard-Jones energy. (a) Total Lennard-Jones energy over time for our $D=20\,$LC, $d=7\,$LC simulation shows an energy barrier. (b) The Lennard-Jones energy barrier increases with increasing grain diameter $D$ and decreasing ideal granule diameter $d$. The shaded surface plot with dashed lines represents the prediction that the energy barrier scales like  $L \approx \sqrt{3}D\left(D/d + 1\right)$, an approximation for the total length of granule boundaries. The black data points are the Lennard-Jones energy barrier heights $U / \epsilon$ as directly computed from grain splitting simulations.
  }
  \label{fig:LennardJones}
\end{figure}

We found that for all simulations, grain splitting causes the dimensionless internal energy $U/\epsilon$ to first increase to a maximum value before ultimately decreasing, just like the entropic term of the free energy. For example, Figure~\ref{fig:LennardJones}a shows one energy barrier from a sample simulation ($d = 7\,$LC and $D=20\,$LC). We find that the height of this Lennard-Jones energy barrier scales like the height of the entropic term of the free energy. It increases with increasing grain diameter $D$ and decreasing ideal granule diameter $d$, following the same scaling prediction $U \propto L \approx \sqrt{3}D\left(D/d + 1\right)$. Therefore, the addition of an interparticle interaction may shift $\Delta F/k_BT$ by an overall constant, but does not change the shape of the free energy landscape.

\bibliography{grainsSplit}